# NEW OPEN CLUSTERS FOUND BY MANUAL MINING OF DATA BASED ON *GAIA* DR2

Juan Casado

Facultad de Ciencias, Universidad Autónoma de Barcelona, 08193, Bellaterra, Catalonia, Spain

Email: juan.casado@uab.cat

*Abstract*

The physical nature of a series of 20 new open clusters is confirmed employing existing data on putative star members, mainly from the second *Gaia* Data Release (DR2). The clusters were discovered as overdensities of stars by visual inspection of either photographic DSS plates or proper motion plots of random source fields. The reported objects are not present in the most comprehensive or recent catalogs of stellar clusters and associations. For all of them, clumps of comoving stars are revealed in the proper motion space. The parallaxes of the clumped stars are compatible with the real existence of open clusters over narrow ranges of distances. Surface density calculations, free of most noise from non-member sources, allow differentiating a cluster core and an extended cluster corona in some instances. Color-magnitude diagrams generally show a definite main sequence that allows the confirmation of the physical existence of the clusters and some of their characteristics. Two of the new clusters seem to form a double system with a common origin. Several of the new clusters challenge the claim of near completeness of the known OC population in the distance range from 1.0 to 1.8 kpc of the Sun (Kharchenko et al. 2013).

## 1. Introduction

Open clusters (OCs) are metastable groups of stars that were born together and still move together as they are gravitationally bound. Nonetheless, OCs slowly dissolve through the ages due to the Galaxy tidal field, interactions between their members, and close encounters with external stars and molecular clouds (Carraro 2006). Consequently, most of the studied OCs are younger than $10^9$ yr (Kharchenko et al. 2013; Liu & Pang 2019). OCs are fundamental laboratories to study the formation and evolution of the stars (e.g., Mermilliod 1981) and the formation and structure of the Galactic disk (Magrini et al. 2017; Cantat-Gaudin et al. 2018). It is currently accepted that most if not all stars form as members of a cluster embedded in its parent molecular cloud (e.g., Portegies Zwart et al. 2010). The key features of OCs to understand stellar evolution are that all members of a given cluster have essentially the same age and initial chemical composition and are at the same distance. Therefore, the different evolutionary stages of these stars depend basically on their masses. For instance, color-magnitude diagrams (CMD) of OCs offer empirical isochrones to compare with the theoretical models of stellar evolution (Marigo et al. 2017; Spada et al. 2017).

From the density of OCs in the solar neighborhood, it can be estimated that the Galaxy contains at least 100,000 open clusters (Koposov et al. 2008). However, no more than 5,000 are known (Bica et al. 2019), and only a small fraction of the Milky Way is well studied (e.g., Kharchenko et al. 2013), which evidences that current Galactic star cluster catalogs are incomplete (Cantat-Gaudin et al. 2019). The main reason is that most stars of the Galaxy cannot be observed with sufficient precision, either due to extinction by the interstellar dust and/or because they are too far away. Nowadays, virtually all new OCs are found through automatic data-mining techniques (e.g., Guo et al. 2018) and sometimes machine-learning algorithms that work out the plethora of data from large stellar databases and new information provided by space missions, such as *Gaia* (Gao et al. 2017; Cantat-Gaudin et al. 2018; Castro-Ginard et al. 2020). The second *Gaia* Data Release (*Gaia* DR2) provides precise full astrometric data (positions, parallax, and proper motions) and three-band photometry for about 1.3 billion stars (Gaia Collaboration 2018). These are powerful tools and, consequently, success in the classical work of individual observers discovering new open clusters by visual observation or inspection of photographic plates is increasingly difficult. However, the number of unknown clusters is so large that there is still some room for this kind of findings. On the other hand, *Gaia* DR2 data allow us to confirm manually if an apparent local overdensity of stars is a real OC or a mere asterism.

The present paper reports a preliminary study using the *Gaia* DR2 catalog of 20 new OC candidates visually detected (Casado 2 to 21; Casado Alessi 1 was previously identified (Casado 1990; Alessi et al. 2003)). The paper is organized as follows. In Section 2, the discovery and first preliminary data of OC candidates are presented, their novelty being verified using up-to-date star cluster catalogs. In Section 3, the proper motions and parallaxes of candidate members are determined and an estimation of distance is provided. In Section 4, the surface densities of possible star members are calculated and used to obtain the OC radius. In Section 5, the CMDs of these new star cluster candidates are presented to assess their real existence. In Section 6, the special case of a new double cluster is described. Some concluding remarks are summarized in Section 7.

## 2. The discovery of new OC candidates.

Some of the present OC candidates were discovered as overdensities of stars by visual inspection of photographic plates closer than 5º to the galactic plane using the Aladin service of the CDS (https://aladin.u-strasbg.fr/AladinLite/), as a part of a long-term survey. For instance, Figure 1 shows the DSS Colored star field around the center of the OC candidate Casado 2, located at $l = 288.62$ and $b = -0.26$ in galactic coordinates ($\alpha = 163.41$ and $\delta = -59.82$ in J2000; determined precisely after studying the OC). A ring-shaped overdensity of not too faint stars (*G*-band magnitude < 18) embedded in nebulosity is perceptible in the central portion of the image.

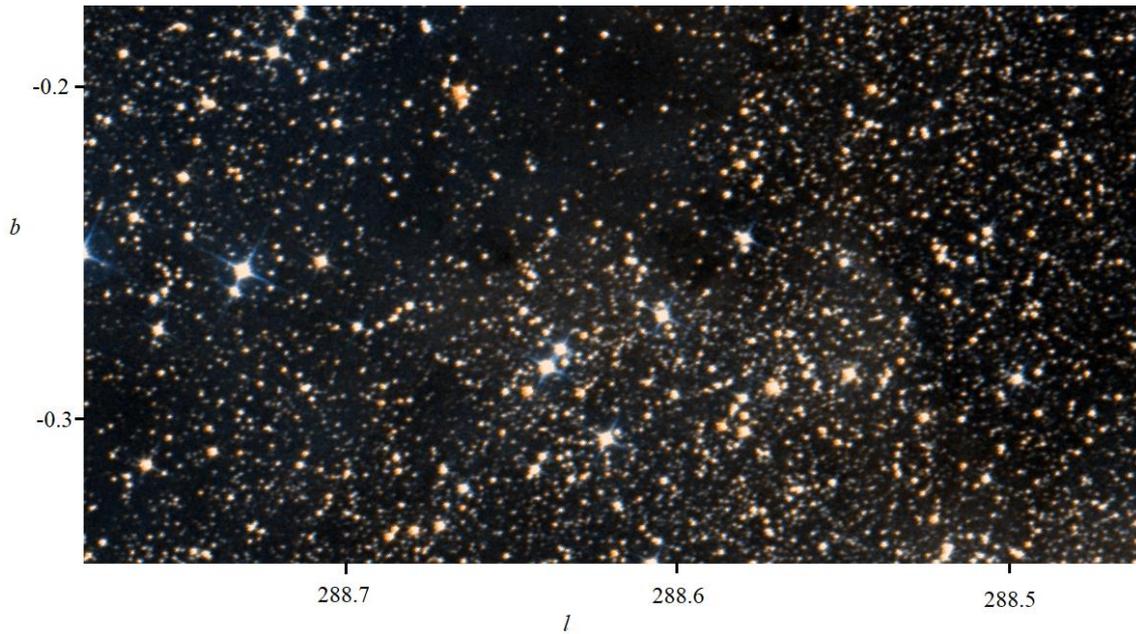

Figure 1. Star field around Casado 2 in the galactic coordinate system. A ring-shaped overdensity of stars near the center of the image was the first hint of the new OC candidate. The horizontal field of view is 19.2 arcmin.

On the other hand, one of the new cluster candidates was first noticed due to a *lower* density of stars in a part of the field (Fig. 2). The anomalous presence of only a few bright stars and bright nebulosity led to a further study of this case. The local density of *Gaia* DR2 sources was also much lower than that of the surrounding field. Such anomaly is due to the obscuration of the background stars by the parent molecular cloud LDN 1670, which is visible in Figure 2 and where new stars were born. At least five young stellar object candidates (YSO) are IR sources in the 2MASS archival image of this young embedded cluster but are not recorded as sources by the *Gaia* mission.

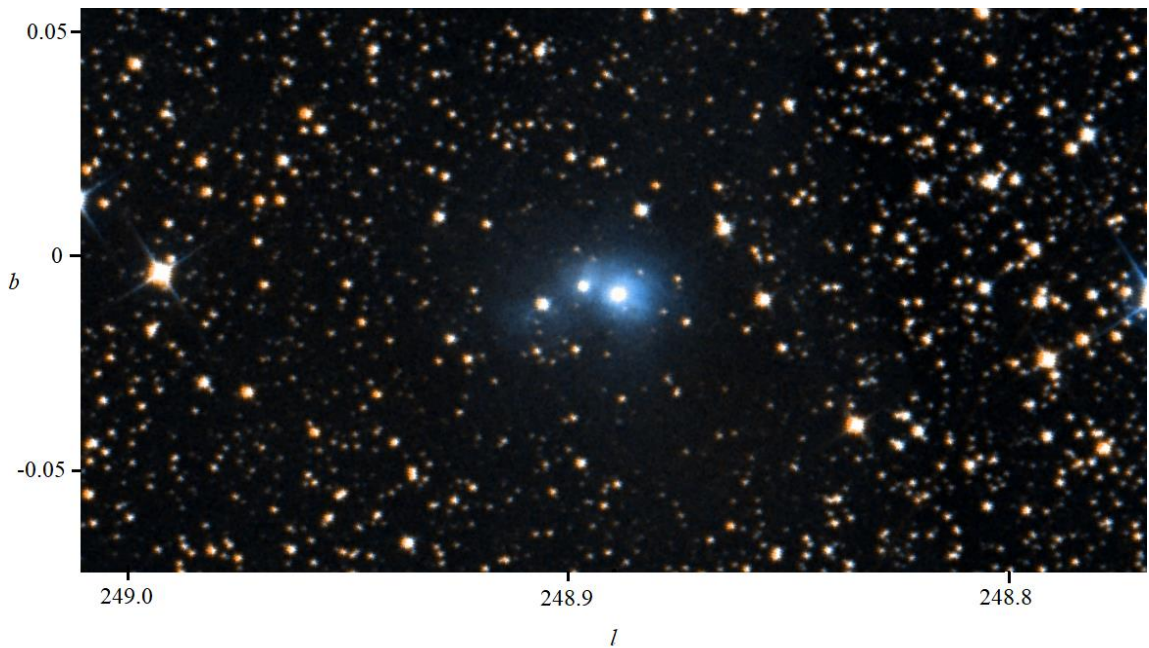

Figure 2. Star field around Casado 3 in the galactic coordinate system. The presence of bright nebulosity and the lower density of stars led to the finding of this OC candidate. The horizontal field of view is 14.5 arcmin.

Finally, most of the OC candidates (entries #4 to #6, #8 to #10, #12 to #16, and #18 to #21 in Table 1) were first found as crowds of sources in the proper motion space of random fields (routinely of 1 x 1 deg$^2$) in selected regions of the galactic disc. For example, Figure 3 shows the plot where Casado 4 was discovered as a distinct patch of sources with consistent proper motions centered at proper motion in α, $\mu_\alpha$ = -5.7 mas yr$^{-1}$; proper motion in δ, $\mu_\delta$ = 5.3 mas yr$^{-1}$. Errors associated with these mean properties are large enough to encompass all the clumped sources and rounded up to one significant figure. In this case, the final ranges of proper motions required as a necessary condition for any star to be considered as a member of Casado 4 are $\mu_\alpha$ = -5.7 ± 0.2 mas yr$^{-1}$, $\mu_\delta$ = 5.3 ± 0.3 mas yr$^{-1}$. The rest of the clumps in the image were not identified as new OCs and show motions similar to the mainstream motions of the field stars, suggesting random crowding of unrelated sources.

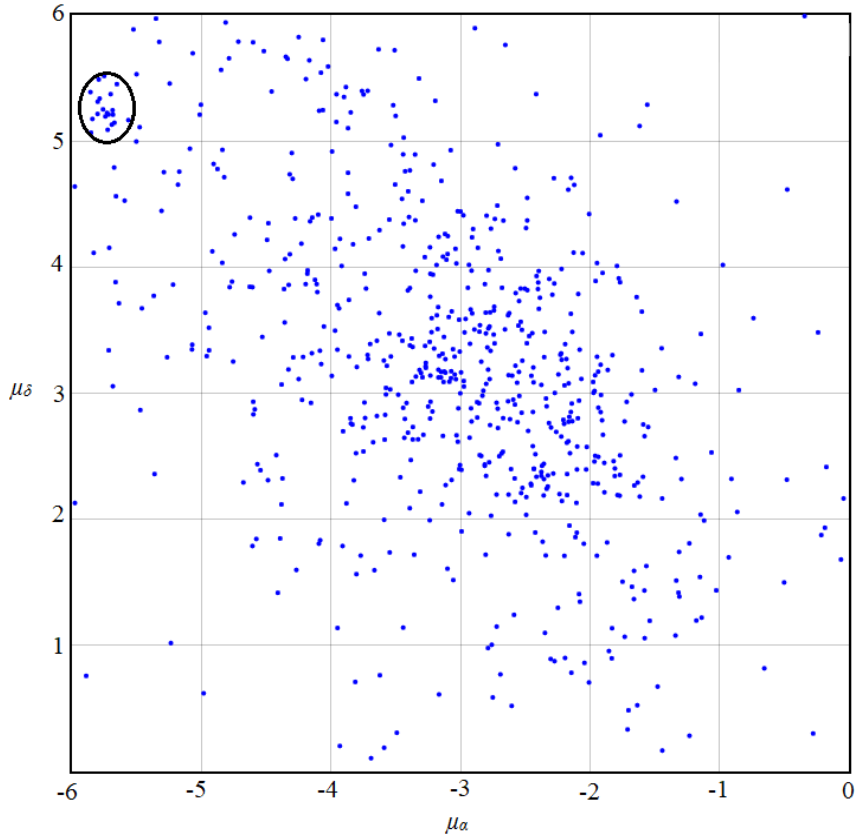

Figure 3. The distribution of proper motions (in mas yr$^{-1}$) of all stars brighter than $G$ = 17 within a 1 x 1 deg$^2$ square centered at α = 125.8 and δ = -37.6. Casado 4 was first detected as a clump of sources at $\mu_\alpha$ = -5.7 ± 0.2 mas yr$^{-1}$, $\mu_\delta$ = 5.3 ± 0.3 mas yr$^{-1}$ (encircled by an ellipse).

All the so detected OC candidates were searched in the most comprehensive and/or recent catalogs of star clusters and similar objects published by the end of May 2020 (e.g., Buckner & Froebrich 2013; Kharchenko et al. 2013; Schmeja et al. 2014; Joshi et al. 2016; Cantat-Gaudin et al. 2018; Bica et al. 2019; Cantat-Gaudin et al. 2019; Liu & Pang 2019; Sim et al. 2019; Castro-Ginard et al. 2020). No known clusters or associations of stars were found which could be compatible with the present candidates; i. e., as a rule, the OC candidates are not located within the angular radius of any of the documented clusters in the literature. Exceptionally, when there is a known OC in the same field of any of the candidates (i.e., the centroids of both objects are closer than the addition of their apparent radii), it has been proven that proper motions and/or parallaxes of both objects are not compatible.

As expected, most of the star overdensities I have found by visual inspection of numerous digitized plates or proper motion plots happened to be already known OCs, embedded clusters, OC remnants, or candidates. Moreover, most of the overdensities not previously cataloged happened to be mere asterisms, i.e., they do not bear the common properties of OCs: parallaxes, proper motions, and coherent CMD. Nonetheless, some of these overdensities are the new OC candidates here reported since their real existence as physical groups can be reasonably proven, as shown in the following Sections.

## 3. Proper motions and parallaxes

Unless otherwise noted, only stars of $G$-band magnitude < 17 have been routinely considered in this study to keep astrometric errors within reasonable limits. Typical parallax errors grow exponentially with magnitude and reach 0.1 mas –which is a substantial fraction of the parallaxes of most of the new clusters (see Table 1)– for stars of magnitude $G \sim 17$ (Fig. 4). Specifically, it has been calculated that for the *Gaia* sources with five-parameter astrometric solutions, the median uncertainty in parallax and position at the reference epoch J2015.5 is about 0.04 mas for bright ($G < 14$) sources, 0.1 mas at $G = 17$, and 0.7 mas at $G = 20$ (Lindegren et al. 2018). The minimum error in proper motions also grows exponentially and exceeds 0.1 mas yr$^{-1}$ around magnitude $G = 17$.

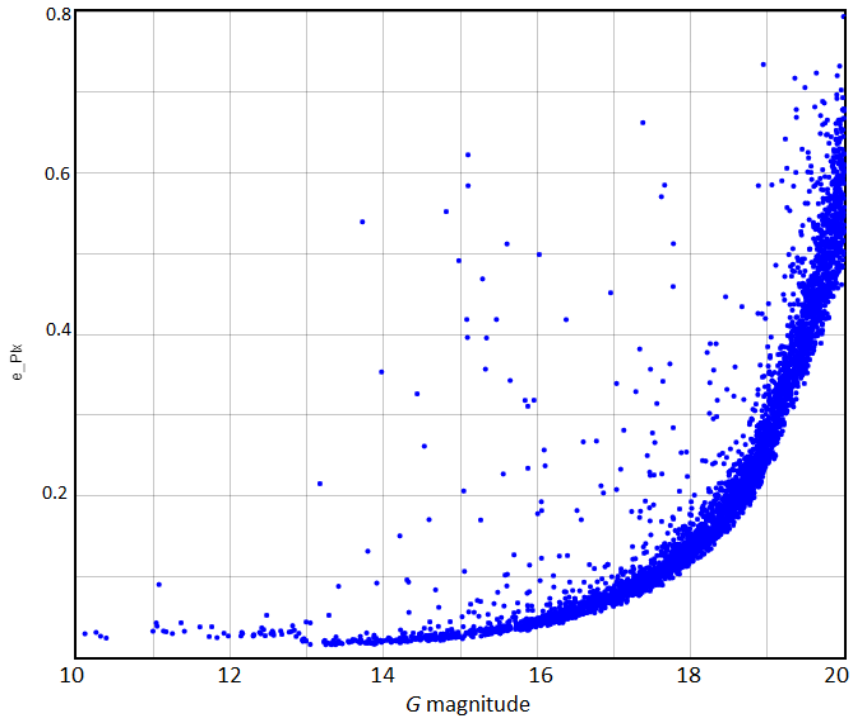

Figure 4. Parallax error (in mas) as a function of $G$-band magnitude for the ensemble of stars in a random field of 100 arcmin.

Once the approximate positions of the subject OC candidates were defined, the proper motions of all of them were determined. For instance, Figure 5 shows the proper motion space of stars brighter than $G = 17$ in a square field of 28 x 28 arcmin$^2$, centered in the OC candidate Casado 5 ($l = 251.20$, $b = -3.15$), which is a disperse grouping, undetectable in the photographic plates, but detached from the field stars in this plot. In this case the proper motions are $\mu_\alpha = -4.7 \pm 0.2$ mas yr$^{-1}$ and $\mu_\delta = 2.2 \pm 0.2$ mas yr$^{-1}$.

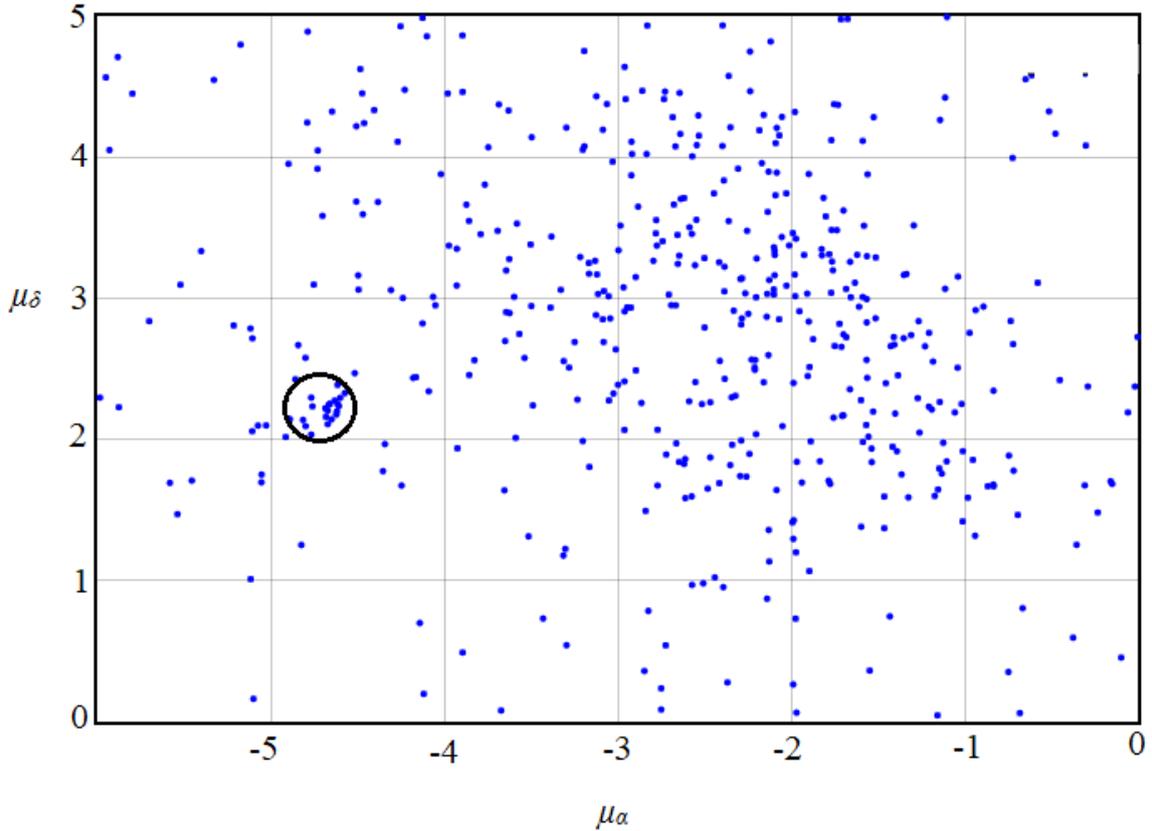

Figure 5. The distribution of proper motions (in mas yr$^{-1}$) of all stars brighter than $G = 17$ within a 28 x 28 arcmin$^2$ square centered at $\alpha = 119.42$ and $\delta = -35.13$. Casado 5 is noticed as a conspicuous clump of sources encircled around $\mu_\alpha = -4.7 \pm 0.2$ mas yr$^{-1}$, $\mu_\delta = 2.2 \pm 0.2$ mas yr$^{-1}$.

Next, the parallaxes of the stars selected according to such proper motions were used to determine the mean parallax of the mentioned clump of comoving stars. Following the same example, the parallaxes of 17 out of 23 comoving stars in a field of radius 14 arcmin are in the range from 0.75 to 0.82 mas, with all individual standard errors lower than 6%. It is thus clear that there is a real concentration of stars in five-dimensional space (i.e., position, proper motions, and parallax), with a mean parallax of $0.79 \pm 0.04$ mas. The median of all these parallaxes is 0.78 mas. A common distance, $d = 1.23 \pm 0.06$ kpc, is inferred after correcting for a global offset of 0.029 mas since *Gaia* DR2 parallaxes are too small on the whole (Lindegren et al. 2018). Nevertheless, it cannot be ruled out that any other star of this field could also be a member of the cluster that is running away as the OC evaporates and which proper motion is different from the mainstream group.

## 4. Surface densities and radii

Surface density $F$ is defined as the number of stars ($N$) per unit area of the celestial sphere. OCs are assumed to have spherical symmetry in surface density calculations, namely,

$$dN = 2\pi r F dr \qquad (1).$$

Surface density profiles are routinely used for cluster radius determination (e.g., Seleznev 2016). A review of arguments in favor of the usual existence of cluster coronae, extended sparse outer regions often beyond the tidal radii of OCs, has been presented by Danilov et al. (2014). The corona radius, i.e., the overall radius of an OC, is defined as the radius where $F$ equals the average stellar density of the surrounding field. Cluster coronae were usually difficult to prove due to low stellar density in the corona and random fluctuations of the stellar density of the background. Therefore, reported OC radii have often been underestimated (Seleznev 2016). Nowadays, however, with the help of source data such as *Gaia* DR2, it is possible to remove most of the contamination due to background stars that are not cluster members, and the presence of OC coronae is more clearly revealed.

Let's see an example. Following the method already described for other OC candidates, Casado 6 (galactic coordinates $l = 249.09$, $b = +0.14$) was determined to have $\mu_\alpha = -2.9 \pm 0.3$ mas yr$^{-1}$, $\mu_\delta = 3.5 \pm 0.3$ mas yr$^{-1}$ and a mean parallax $plx = 0.22 \pm 0.04$ mas. To get rid of the stars with different proper motions and parallax (which most probably are non-members of the OC), in this Section, only the sources constrained by the quoted proper motions, both in $\alpha$ and $\delta$, and parallax are considered. In this way, most noise from non-member field sources is removed. The positions of the selected stars in a wide field of 60 arcmin are plotted in Figure 6. We may notice a concentration of sources near the center of the plot, the OC core ($r <$ 1arcmin), and an apparent cluster corona extending to $r \sim 4$ arcmin.

These visual estimations have been refined through surface density analysis. $F$ is maximal at the inner core (ca. 3 arcmin$^{-2}$). Then, the surface density profile decreases gradually down to practically coincide with the background surface density for concentric annuli of $r = 5$ arcmin and wider. These results allow estimating an OC actual radius $R$ of $5 \pm 1$ arcmin, which is significantly more extended than the cluster core (Fig. 6). From an estimated parallax distance of 4.0 kpc, the physical radius of the cluster would be around 5 pc, which is normal for OCs.

A preliminary number of plausible star members is established by counting the *Gaia* sources within such radius that fulfill the constraining conditions of the OC; i.e., only the stars having position, parallax, and proper motions within the constraining ranges so established are selected as probable OC members. In this case, the number of putative cluster members ($G < 17$) with $r < 5$ arcmin is 21, which results in an average surface density $F = 0.27$ arcmin$^{-2}$, ca. 5 times the estimated background density. Such a large ratio of average surface density members versus background is obtained due to the noise removal method used, which is remarkable since this cluster (as most of the OCs here described) is barely noticeable in the DSS plates.

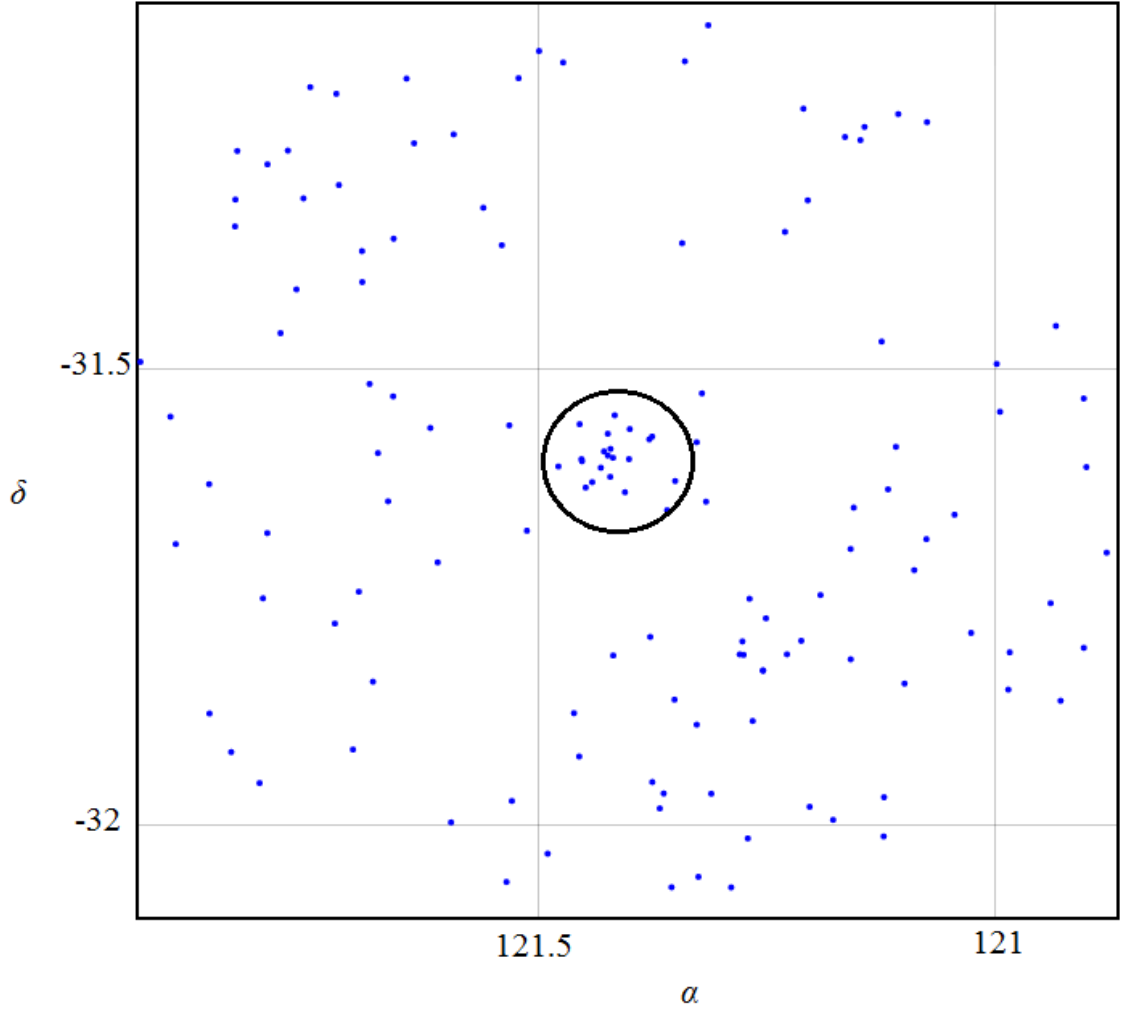

Figure 6. Location of potential members of OC Casado 6 in J2000 coordinates. Most noise from non-member field sources has been removed, and both the OC core and corona are noticeable. A circle encloses the most probable OC members.

## 5. Color-Magnitude Diagrams

Finally, the CMD diagrams were obtained to ratify the existence of the new OC and to confirm or modify some of the basic data found, such as the distance and the number of member stars. Let's consider, as an example, the small OC Casado 7 (Table 1), which is barely detectable as an overdensity of stars apparently near the molecular cloud PGCC G000.18-00.50, which is however much farther away (3.24 kpc; Ade et al. 2016). To select only the most probable members of the cluster, the CMD was constructed with the stars of $r < 2.5$ arcmin that fulfill the constraining conditions: $plx = 0.58 \pm 0.06$ mas, $\mu_\alpha = 0.5 \pm 0.5$ mas yr$^{-1}$, and $\mu_\delta = -1.5 \pm 0.5$ mas yr$^{-1}$. In this case, dimmer stars ($G < 18$) were also used to complete the tail of the main sequence (this procedure has been applied in other CMD, as reported in Table 1). In spite that the total number of sources with such characteristics is only 21, the resulting CMD (Fig. 7) shows a defined main sequence.

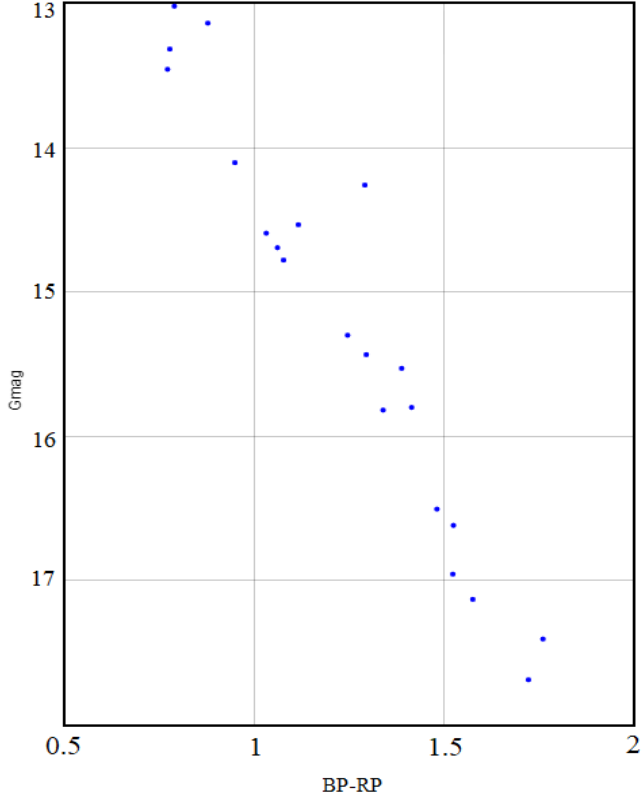

Figure 7. *G* magnitude versus color index BP-RP, which refers to the difference between *Gaia* blue and red magnitudes, for the most probable members of OC Casado 7.

The temperature estimate of the bluest star in Figure 7 (*Gaia* DR2 source 4057305315042854144) varies in the literature between $T_{eff}$ ~ 6,000 K (*Gaia* data estimate from Apsis-Priam) and $T_{eff}$ ~ 11,000 K in the VVV survey (Grosbøl & Carraro 2018). This could be due to the existence of a close companion less than 4 arcsec away, which may contaminate the source data. On the other hand, the second bluest *Gaia* DR2 source (4057305353733574400) is well resolved. It is most probably a main-sequence star, taking into account its position in Figure 7 (in the upper-left region of the main sequence) and the fact that it is considered a solar-type dwarf (Nascimbeni et al. 2016). Reported data for this source distance (*d*) range from 1.56 kpc (Anders et al. 2019) to 1.63 kpc (Bailer-Jones et al. 2018), confirming the OC distance (1.6 kpc) obtained here from its median parallax. The estimated stellar mass according to the StarHorse catalog (Anders et al. 2019) is 2.3 ± 0.5 solar masses. Then, assuming that this is the closest star to the turn-off point of the CMD towards the red giant branch, Equation (2) can be applied to estimate the time ($t_{ms}$ in yr) such a star populates the main sequence (Hansen et al. 2012), which should coincide with the approximate age of the OC:

$$t_{ms} = 10^{10} (M_{star})^{-2.5} \qquad (2).$$

For $M_{star}$ = 2.3 ± 0.5 solar masses, log $t_{ms}$ = 9.1 ± 0.3, which is, though rough, a preliminary estimate of the OC age.

Let's consider another example: Figure 8 shows the CMD diagram of Casado 8 (Table 1). It contains all the sources of *G* < 17, *r* < 14 arcmin, *plx* = 0.28 ± 0.06 mas , $\mu_\alpha$ = -3.0 ± 0.2   mas

yr$^{-1}$, and $\mu_\delta$ = 1.7 ± 0.5 mas yr$^{-1}$. In this case, I have been less restrictive to include all putative member stars within the OC corona, even if some outliers are encompassed in the graph. As a consequence, the main-sequence band appears broadened, especially at the lower end of fainter stars. However, the left-hand side of the main sequence, from whose stars we expect to obtain the most relevant data, remains well-defined.

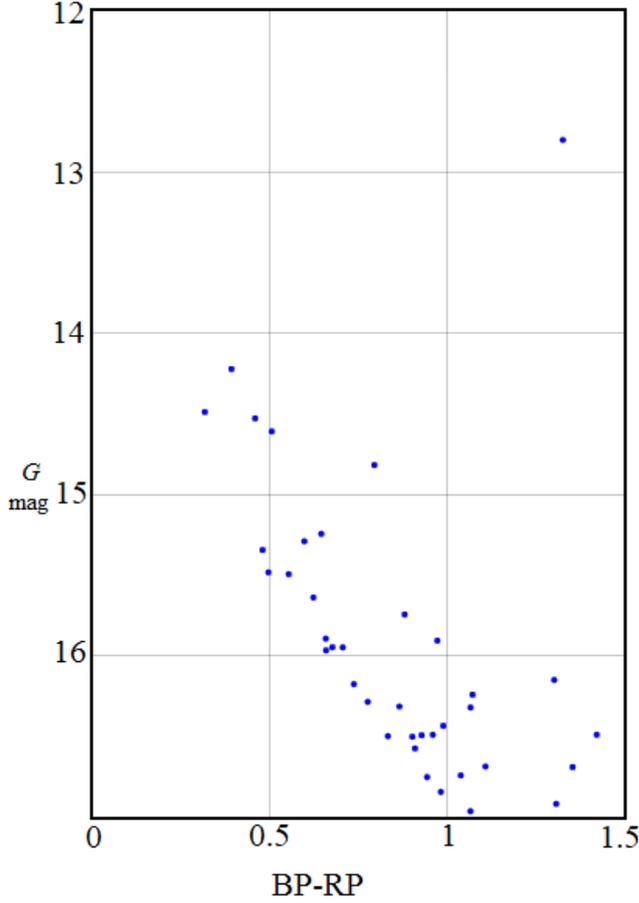

Figure 8. *G* magnitude versus color index BP-RP for the putative members of OC Casado 8.

The brighter star in this diagram (*Gaia* DR2 source 5548313485403766016) is certainly far away from the main sequence. Accordingly, it is classified as a giant star in the TESS Input Catalog (Stassun et al. 2019), and its position in the CMD confirms it as a red giant. Its reported distances range from 2.6 ± 0.2 kpc (Anders et al. 2019) to 2.8 ± 0.3 kpc (Bailer-Jones et al. 2018), which are compatible with the parallax distance obtained for the whole OC (Table 1).

The bluest star that seems to lie in the main sequence (*Gaia* DR2 source 5548316547730573056) would most probably be, however, an evolved star (Gaia DR2 x AllWise catalog, Marton et al. 2019), despite being also classified as a dwarf (Stassun et al. 2019). In any case, all reported distances are 2.8 ± 0.5 kpc (Bailer-Jones et al. 2018; Anders et al. 2019), compatible with the OC distance obtained in this work (Table 1). The same classifications are assigned to the next brighter star on the left-hand side of the main sequence (*Gaia* DR2 source 5548367056548165504). Anyway, the reported distances of 3.4 ± 0.9 kpc

(Anders et al. 2019) and 3.6 kpc ± 18% (Bailer-Jones et al. 2018) are also well-matched with the presently estimated distance of the new OC.

Similar analyses have been applied to other OC candidates and their stars. It is noteworthy that a red giant (*Gaia* DR2 source 5545561893139329664; Stassun et al. 2019) of spectral type M0.3V (Pickles & Depagne 2010) was also found in Casado 5. Consistently reported distances of that star, ranging from 1.17 kpc (Anders et al. 2019) to 1.19 kpc (Bailer-Jones et al. 2018), agree on the distance found for the new OC. By the way, a distance as low as 27 pc was also reported for the same star (Pickles & Depagne 2010). Casado 14 also includes a red giant (*Gaia* DR2 source 4277812208875222560; Stassun et al. 2019). Recently reported distances of that star are 3.9 ± 0.6 kpc (Bailer-Jones et al. 2018) and 4.0 ± 0.7 kpc (Anders et al. 2019), also matching the parent OC distance (Table 1).

An iterative analysis, refining the membership and properties of every OC candidate from first estimates, was repeated until achieving self-consistency and accuracy of them. All these studies lead to the confirmation of the new OCs as genuine physical systems with reliable data that are summarized in Table 1.

Table 1. Main results for 20 new OCs obtained in this study from *Gaia* DR2 data for stars brighter than $G = 17$ (unless otherwise noted) and constrained by consistent proper motions and parallaxes (see text for details). Positions are in galactic coordinates. *N* refers to the approximate number of member stars within the adopted radius *R*.

| OC # | *l* degree | *b* degree | *plx* mas | *d* kpc | $\mu_\alpha$ mas yr$^{-1}$ | $\mu_\delta$ mas yr$^{-1}$ | *R* arcmin | *N* |
|---|---|---|---|---|---|---|---|---|
| 2  | 288.62 | -0.26 | 0.39 ± 0.07 | 2.4 ± 0.5  | -6.9 ± 0.2 | 2.6 ± 0.2  | 2 ± 0.5   | 15 |
| 3  | 248.90 | -0.01 | 0.83 ± 0.09 | 1.16 ± 0.1 | -4.7 ± 0.7 | 4.8 ± 0.7  | 4 ± 0.5   | 15[a,b] |
| 4  | 256.10 | -0.21 | 0.81 ± 0.08 | 1.2 ± 0.1  | -5.7 ± 0.2 | 5.3 ± 0.3  | 10 ± 1    | 20 |
| 5  | 251.20 | -3.15 | 0.78 ± 0.04 | 1.23 ± 0.06| -4.7 ± 0.2 | 2.2 ± 0.2  | 14 ± 1    | 20 |
| 6  | 249.09 | 0.14  | 0.22 ± 0.04 | 4.0 ± 0.7  | -2.9 ± 0.3 | 3.5 ± 0.3  | 5 ± 1     | 24[a] |
| 7  | 0.24   | -0.50 | 0.58 ± 0.06 | 1.6 ± 0.2  | 0.5 ± 0.5  | -1.5 ± 0.5 | 2.5 ± 0.5 | 21[a] |
| 8  | 249.51 | 1.95  | 0.28 ± 0.06 | 3.3 ± 0.7  | -3.0 ± 0.2 | 1.7 ± 0.5  | 10 ± 1    | 30 |
| 9  | 249.33 | -0.69 | 0.24 ± 0.05 | 3.7 ± 0.6  | -3.2 ± 0.3 | 3.7 ± 0.2  | 5 ± 1     | 25[a] |
| 10 | 249.16 | -0.75 | 0.25 ± 0.05 | 3.6 ± 0.6  | -3.1 ± 0.3 | 3.6 ± 0.2  | 3.5 ± 0.5 | 27[a] |
| 11 | 119.86 | -6.04 | 0.59 ± 0.06 | 1.6 ± 0.2  | -3.3 ± 0.2 | -2.6 ± 0.3 | 6 ± 1     | 11 |
| 12 | 122.89 | 0.78  | 0.34 ± 0.08 | 2.8 ± 0.6  | -3.0 ± 0.2 | -0.4 ± 0.2 | 3.5 ± 0.5 | 15[a] |
| 13 | 121.80 | 0.73  | 0.26 ± 0.04 | 3.4 ± 0.4  | -1.7 ± 0.3 | -0.3 ± 0.3 | 5 ± 1     | 21[a] |
| 14 | 122.04 | -0.26 | 0.21 ± 0.06 | 4.2 ± 1    | -2.1 ± 0.2 | -0.4 ± 0.2 | 3 ± 0.5   | 26[c] |
| 15 | 118.51 | -0.08 | 0.34 ± 0.06 | 2.8 ± 0.5  | -2.1 ± 0.2 | -0.8 ± 0.2 | 4.5 ± 0.5 | 20[a] |
| 16 | 117.49 | 0.16  | 0.36 ± 0.09 | 2.6 ± 0.6  | -2.1 ± 0.3 | -0.9 ± 0.3 | 4 ± 0.5   | 33[d] |
| 17 | 118.90 | 0.97  | 0.47 ± 0.07 | 2.0 ± 0.3  | -3.3 ± 0.3 | -1.0 ± 0.2 | 4 ± 0.5   | 20 |
| 18 | 113.88 | 0.35  | 0.37 ± 0.1  | 2,5 ± 0.8  | -3.7 ± 0.2 | -1.6 ± 0.3 | 4 ± 0.5   | 21[d] |
| 19 | 108.93 | -2.01 | 0.30 ± 0.08 | 3.1 ± 0.7  | -3.3 ± 0.3 | -2.2 ± 0.3 | 3 ± 0.5   | 17[d] |
| 20 | 109.63 | -3.46 | 0.52 ± 0.06 | 1.8 ± 0.2  | -3.3 ± 0.1 | -2.6 ± 0.2 | 3 ± 0.5   | 12 |
| 21 | 107.77 | 0.26  | 0.38 ± 0.09 | 2.4 ± 0.4  | -2.7 ± 0.2 | -1.9 ± 0.2 | 7 ± 1     | 30[d] |

[a] $G < 18$; [b] Including YSO candidates; [c] $G < 20$; [d] $G < 19$

The ensemble of data in Table 1 shows some trends. Most of the new clusters are relatively distant (*plx* < 0.9 mas), small (11 to 33 stars), and disperse to the point of being indistinguishable from the field stars. These features suggest that automatic surveys of OCs could overlook, at least in part, this class of objects. In any case, the derived physical dimensions of the OCs span from ~3 pc (Casado 7) to ~19 pc (Casado 8), matching the general statistics of OCs. Fifteen of the new OCs are within 1° of the galactic plane, and 19 of them are closer than 3.5° to the galactic plane. The typical spread in mean parallaxes is ~ 0.06 mas and in proper motions is ~ 0.3 mas yr$^{-1}$. Thus, there are practical limits for the application of the astrometric *Gaia* data to this kind of study, which approximately are $G < 18$ and $plx > 0.20$ mas.

## 6. A new double cluster

Two of the clusters in Table 1, namely Casado 9 and 10, are very close in position and very similar in proper motions and parallaxes. However, they are not a single spread cluster but a rather conspicuous double cluster (Fig. 9). Two OC cores with a common corona spanning almost 30 arcmin are visible. In this case, the spherical symmetry used for surface density calculations is not applicable.

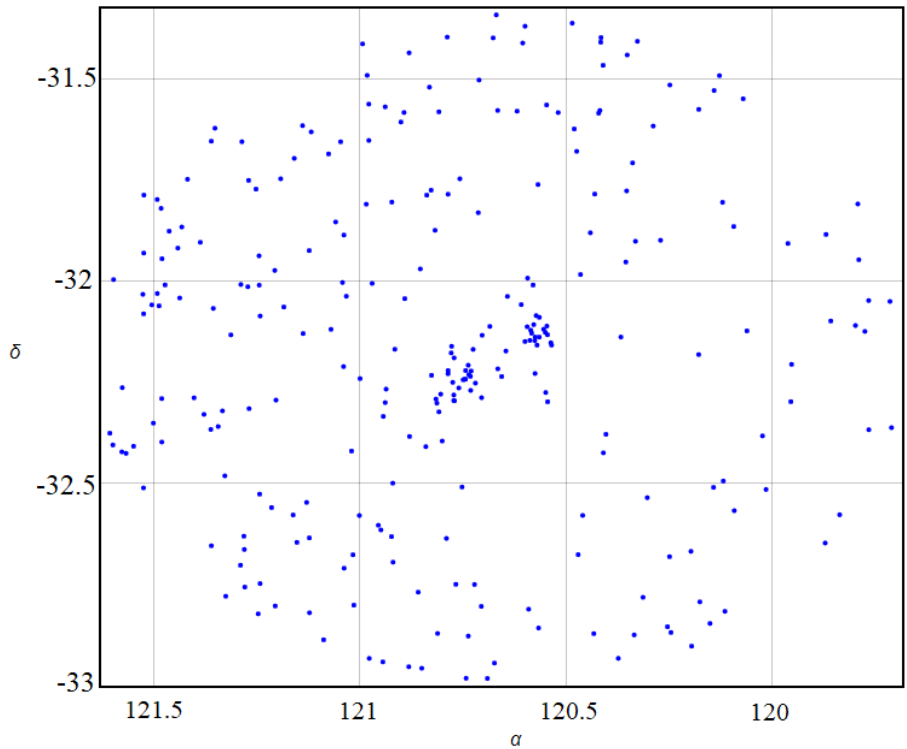

Figure 9. Location of putative members of OCs Casado 9 and 10 in J2000 coordinates. Most noise from non-member field sources is removed, and both OCs are discernible.

To test the hypothesis of the double cluster, the overall CMD diagram of the whole system (Fig. 10) has been built with all the sources of $G < 18$ that fulfill the constraints of both OC (Table 1). Although some spreading of stars is noticed, both main sequences fit perfectly, and the distances reported for individual stars are consistent with those reported in Table 1.

Again, some discrepancies are found in the reported data for individual stars. The brightest source (*Gaia* DR2 5595803217986212736) is classified either as a main-sequence star (Nascimbeni et al. 2016; Marton et al. 2019) or a giant star (Stassun et al. 2019). It could be a blue straggler. In any case, the reported distances to this source are $3.2 \pm 0.4$ kpc (Anders et al. 2019) and $3.6 \pm 0.4$ kpc (Bailer-Jones et al. 2018) and are compatible with the estimated distances of both clusters (Table 1).

On the other hand, the second bluest star in Figure 10 (*Gaia* DR2 5595755251789398656) is classified as a dwarf by Stassun et al. (2019). However, it is unclear if it is the closest star to the turn-off point of the CMD. Its reported mass ranges from 1.76 solar masses (Stassun et al. 2019) to 2.70 solar masses (Anders et al. 2019). The large uncertainty in turn-off point and stellar mass precludes the determination of a reliable age for this system. Fortunately, the reported distances of $3.4 \pm 0.5$ kpc (Anders et al. 2019) and $3.7 \pm 0.3$ kpc (Bailer-Jones et al. 2018) are again compatible with our distance estimate (Table 1). On the whole, these results indicate that all the member stars of this double cluster most probably have a common origin.

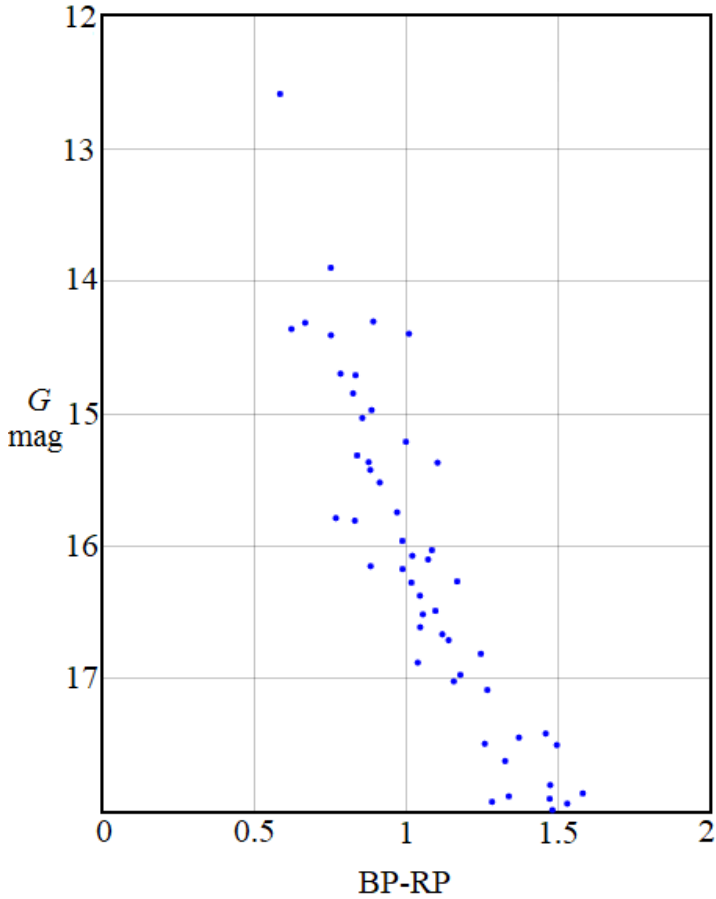

Figure 10. Distribution of the most probable members of the double cluster comprising Casado 9 and 10 in the *G* magnitude versus BP-RP color plane.

## 7. Concluding remarks

The discovery and the first study, mainly based on *Gaia* DR2 data, of twenty new OCs are reported. The OC candidates were found either as a slight overdensity of stars in DSS plates or as a clump of comoving stars in the proper motion space of random regions of the sky near the galactic plane. They are not present in the most comprehensive catalogs of such celestial objects published by the end of May 2020.

The ensemble of comoving stars allows establishing mean parallaxes and the corresponding distances of the new OCs, which are in the range from 1.2 to 4.2 kpc (Table 1).

*Gaia* DR2 data allow removing most of the background star noise. Subsequent surface density analysis confirmed in some cases the presence of an OC core and an extended corona, which size leads to the actual radii of the clusters *R*. The derived physical dimensions match the known OC population.

The CMD diagrams show a definite main sequence for the new OCs. Three of them most probably contain a red giant.

All the new OCs have few stars (11 to 33) and are disperse. The automatic surveys of OCs could overlook this class of objects. Otherwise, they probably would have been detected earlier.

One of the new OCs (Casado 3) is being born, judging by the presence of several YSO candidates and an associated molecular cloud. Two of the OCs (Casado 9 and 10) appear to form a double cluster with compatible properties but with two separate concentrations of stars. Thus, all such stars probably share a common origin.

The claim of near-completeness of the known OC population in the *d* range from 1.0 to 1.8 kpc (Kharchenko et al. 2013) is questioned as six out of 20 new OCs lie precisely in such interval of *d* (see also Cantat-Gaudin et al. 2019).

Finally, it is shown that new OCs can be found *manually*. *Gaia* DR2 data allow their study and eventual confirmation as real physical systems. Following this method, a limited number of researchers could significantly contribute to increasing the number of known OCs.

## Acknowledgments

This work has made use of data from the European Space Agency (ESA) mission Gaia (https://www.cosmos.esa.int/gaia), processed by the Gaia Data Processing and Analysis Consortium (DPAC, https://www.cosmos.esa.int/web/gaia/dpac/consortium). Funding for the DPAC has been provided by national institutions, in particular the institutions participating in the Gaia Multilateral Agreement. This research has made use of the VizieR catalogue access tool, CDS, Strasbourg, France (DOI: 10.26093/cds/vizier).

*References*

Ade, P. A. R., Aghanim, N., Arnaud, M., Ashdown, M., Aumont, J., Baccigalupi, C., ... & Benabed, K. (2016). Planck 2015 results-XXVIII. The Planck Catalogue of Galactic cold clumps. *A&A*, *594*, A28.

Alessi, B. S., Moitinho, A., & Dias, W. S. (2003). Searching for unknown open clusters in the Tycho-2 catalog. *A&A*, *410*(2), 565-575.

Anders, F., Khalatyan, A., Chiappini, C., Queiroz, A. B., Santiago, B. X., Jordi, C., ... & Cantat-Gaudin, T. (2019). Photo-astrometric distances, extinctions, and astrophysical parameters for *Gaia* DR2 stars brighter than G = 18. *A&A*, *628*, A94.

Bailer-Jones, C. A. L., Rybizki, J., Fouesneau, M., Mantelet, G., & Andrae, R. (2018). Estimating distance from parallaxes. IV. Distances to 1.33 billion stars in Gaia data release 2. *AJ*, *156*(2), 58.

Bica, E., Pavani, D. B., Bonatto, C. J., & Lima, E. F. (2019). A Multi-band Catalog of 10978 Star Clusters, Associations, and Candidates in the Milky Way. *AJ*, *157*(1), 12.

Buckner, A. S., & Froebrich, D. (2013). Properties of star clusters–I. Automatic distance and extinction estimates. *MNRAS*, *436*(2), 1465-1478.

Cantat-Gaudin, T., Jordi, C., Vallenari, A., Bragaglia, A., Balaguer-Núñez, L., Soubiran, C., ... & Casamiquela, L. (2018). A *Gaia* DR2 view of the open cluster population in the Milky Way. *A&A*, *618*, A93.

Cantat-Gaudin, T., Krone-Martins, A., Sedaghat, N., Farahi, A., de Souza, R. S., Skalidis, R., ... & Moitinho, A. (2019). *Gaia* DR2 unravels incompleteness of nearby cluster population: new open clusters in the direction of Perseus. *A&A*, *624*, A126.

Carraro, G. (2006). Open cluster remnants: an observational overview. *Bulletin of the Astronomical Society of India*, *34*, 153.

Casado, J. (1990). Fotometría fotográfica de un posible cúmulo abierto. *Astrum*, *Boletín de la Agrupación Astronómica de Sabadell, 91*, 12.

Castro-Ginard, A., Jordi, C., Luri, X., Julbe, F., Morvan, M., Balaguer-Núñez, L., & Cantat-Gaudin, T. (2018). A new method for unveiling open clusters in Gaia-New nearby open clusters confirmed by DR2. *A&A*, *618*, A59.

Castro-Ginard, A., Jordi, C., Luri, X., Cid-Fuentes, J. Á., Casamiquela, L., Anders, F., ... & Badia, R. M. (2020). Hunting for open clusters in Gaia DR2: 582 new open clusters in the Galactic disc. *A&A*, *635*, A45.

Danilov, V. M., Putkov, S. I., & Seleznev, A. F. (2014). Dynamics of the coronas of open star clusters. *Astronomy Reports*, *58*(12), 906-921.

Gaia Collaboration, Brown, A. G. A., Vallenari, A., Prusti, T., de Bruijne, J. H. J., Babusiaux, C., ... & Plachy, E. (2018). Gaia Data Release 2 Summary of the contents and survey properties. *A&A*, *616*(1).

Gao, X. H. (2017). An application of the KNND method for detecting nearby open clusters based on Gaia-DR1. *RAA*, *17*(6), 058.

Grosbøl, P., & Carraro, G. (2018). The spiral potential of the Milky Way. *A&A*, *619*, A50.

Guo, J. C., Zhang, H. W., Zhang, H. H., Liu, X. W., Yuan, H. B., Huang, Y., ... & Chen, B. Q. (2018). New open cluster candidates discovered in the XSTPS-GAC survey. *RAA*, *18*(3), 032.

Hansen, C. J., Kawaler, S. D., & Trimble, V. (2012). *Stellar interiors: physical principles, structure, and evolution*. Springer Science & Business Media. ISBN 978-0-387-94138-7.

Joshi, Y. C., Dambis, A. K., Pandey, A. K., & Joshi, S. (2016). VizieR Online Data Catalog: Open clusters within 1.8 kpc of the Sun (Joshi+, 2016). *VizieR Online Data Catalog*, *359*.

Kharchenko, N. V., Piskunov, A. E., Schilbach, E., Röser, S., & Scholz, R. D. (2013). Global survey of star clusters in the Milky Way-II. The catalogue of basic parameters. *A&A*, *558*, A53.

Koposov, S. E., Glushkova, E. V., & Zolotukhin, I. Y. (2008). Automated search for Galactic star clusters in large multiband surveys-I. Discovery of 15 new open clusters in the Galactic anticenter region. *A&A*, *486*(3), 771-777.

Lindegren, L., Hernández, J., Bombrun, A., Klioner, S., Bastian, U., Ramos-Lerate, M., ... & Lammers, U. (2018). Gaia Data Release 2-The astrometric solution. *A&A*, *616*, A2.

Liu, L., & Pang, X. (2019). A catalog of newly identified star clusters in GAIA DR2. *ApJS*, *245*(2), 32.

Magrini, L., Randich, S., Kordopatis, G., Prantzos, N., Romano, D., Chieffi, A., ... & Bragaglia, A. (2017). The Gaia-ESO Survey: radial distribution of abundances in the Galactic disc from open clusters and young-field stars. *A&A*, *603*, A2.

Marigo, P., Girardi, L., Bressan, A., Rosenfield, P., Aringer, B., Chen, Y., ... & Trabucchi, M. (2017). A new generation of PARSEC-COLIBRI stellar isochrones including the TP-AGB phase. *ApJ*, *835*(1), 77.

Marton, G., Ábrahám, P., Szegedi-Elek, E., Varga, J., Kun, M., Kóspál, Á., ... & Kiss, C. (2019). Identification of Young Stellar Object candidates in the Gaia DR2 x AllWISE catalogue with machine learning methods. *MNRAS*, *487*(2), 2522-2537.

Mermilliod, J. C. (1981). Comparative studies of young open clusters. III-Empirical isochronous curves and the zero age main sequence. *A&A*, *97*, 235-244.

Nascimbeni, V., Piotto, G., Ortolani, S., Giuffrida, G., Marrese, P. M., Magrin, D., ... & Pollacco, D. (2016). An all-sky catalogue of solar-type dwarfs for exoplanetary transit surveys. *MNRAS*, *463*(4), 4210-4222.

Pickles, A., & Depagne, É. (2010). All-Sky Spectrally Matched UBVRI-ZY and u′ g′ r′ i′ z′ Magnitudes for Stars in the Tycho2 Catalog. *PASP*, *122*(898), 1437.

Portegies Zwart, S. F., McMillan, S. L., & Gieles, M. (2010). Young massive star clusters. *ARA&A*, *48*, 431-493.

Schmeja, S., Kharchenko, N. V., Piskunov, A. E., Röser, S., Schilbach, E., Froebrich, D., & Scholz, R. D. (2014). VizieR Online Data Catalog: Milky Way global survey of star clusters. III.(Schmeja+, 2014). *VizieR Online Data Catalog*, *356*.

Seleznev, A. F. (2016). Open-cluster density profiles derived using a kernel estimator. *MNRAS*, *456*(4), 3757-3773.

Sim, G., Lee, S. H., Ann, H. B., & Kim, S. (2019). 207 NEW OPEN STAR CLUSTERS WITHIN 1 KPC FROM GAIA DATA RELEASE 2. *Journal of the Korean Astronomical Society*, *52*(5), 145-158.

Spada, F., Demarque, P., Kim, Y. C., Boyajian, T. S., & Brewer, J. M. (2017). The Yale–Potsdam Stellar Isochrones. *ApJ*, *838*(2), 161.

Stassun, K. G., Oelkers, R. J., Paegert, M., Torres, G., Pepper, J., De Lee, N., ... & Rojas-Ayala, B. (2019). The Revised TESS Input Catalog and Candidate Target List. *AJ*, *158*(4), 138.